\documentstyle[epsf,a4]{article}  
\begin{document}                
\newcommand{\manual}{rm}        
\newcommand\bs{\char '134 }     
\newcommand{\Het}{$^3{\mathrm{He}}$}
\newcommand{\Hef}{$^4{\mathrm{He}}$}
\newcommand{\A}{{\mathrm{A}}}
\newcommand{\D}{{\mathrm{D}}}
\newcommand{\simlt}{\stackrel{<}{{}_\sim}}
\newcommand{\simgt}{\stackrel{>}{{}_\sim}}
\newcommand{\MeV}{\;\mathrm{MeV}}
\newcommand{\TeV}{\;\mathrm{TeV}}
\newcommand{\GeV}{\;\mathrm{GeV}}
\newcommand{\eV}{\;\mathrm{eV}}
\newcommand{\cm}{\;\mathrm{cm}}
\newcommand{\s}{\;\mathrm{s}}
\newcommand{\sr}{\;\mathrm{sr}}
\newcommand{\lab}{\mathrm{lab}}
\newcommand{\ts}{\textstyle}
\newcommand{\ol}{\overline}
\newcommand{\be}{\begin{equation}}
\newcommand{\ee}{\end{equation}}
\newcommand{\ba}{\begin{eqnarray}}
\newcommand{\ea}{\end{eqnarray}}
\newcommand{\rau}{\rho_{\mathrm Au}}
\newcommand{\nn}{\nonumber}
\newcommand{\nm}{{\nu_\mu}}
\newcommand{\st}{\sin^2\theta_{\mathrm W}}
\newcommand{\N}{{\mathrm{N}}}
\newcommand{\css}{({\mathrm{cm}}^2-{\mathrm{s}}-{\mathrm{sr}})^{-1}}
\newcommand{\pp}{$\overline{p}(p)-p\;\;$}
\renewcommand{\floatpagefraction}{1.}
\renewcommand{\topfraction}{1.}
\renewcommand{\bottomfraction}{1.}
\renewcommand{\textfraction}{0.}               
\renewcommand{\thefootnote}{F\arabic{footnote}}
\title{Effect of possible stronger neutrino interaction at
$E_\nm \sim 10^{11}\GeV$ upon the extraction of $\st$ from
the neutral-current cross section at NuTeV}
\author{Saul Barshay and Georg Kreyerhoff\\
III. Physikalisches Institut\\
RWTH Aachen\\D-52056 Aachen\\Germany}
\maketitle
\begin{abstract}                
The possibility exists that cosmic-ray neutrinos with energies
of $\sim 10^{20}\eV$ interact in the atmosphere with a 
cross section at the millibarn level, giving rise to some of the highest-energy air
showers. In a specific dynamical model which can give rise
to such a cross section, we show that there can be a small effect
upon the extraction of the effective value of $\st$ from
$\nm$-hadron, neutral-current cross section data at NuTeV.   
\end{abstract}
The possibility that neutrinos, with energies of the order of $10^{20}\eV$,
have stronger interactions with hadrons than those which are given by
electroweak theory, has been considered, specifically in connection with the 
puzzling observation of the highest energy cosmic-ray air shower events.\cite{ref1,ref2}
Such energetic neutrinos can propagate from sources at cosmological distances
(red shifts $\simgt 1$), undisturbed by the cosmic microwave background radiation,
which photons largely prevent protons of $\simgt 0.6\times 10^{20}\eV$ from
arriving here.\cite{ref3,ref4} Recently\cite{ref5} we have explicitly calculated
a $\sigma_\nm$ of $\sim 10^{-29}\cm^2$ at $E_\nm\sim 10^{20}\eV$, in a specific
dynamical model which involves the hypothesis of a strong, parity-conserving interaction
of $\nm$ with a hypothetical, point-like component of a $\pi^0$, leading to a very
massive neutral lepton $L$, with $m_L\sim 2\times 10^6\GeV$.\footnote{
The possibility of an analogous hypothetical structure (with a charged pion)
for the muon, can lead to a positive-definite, unusual contribution to 
$(g_\mu-2)/2$ at the level of $10^{-9}$, for this value of $m_L$.\cite{ref5,ref6}
However, in addition to the squared coupling, 
the magnitude is rather sensitive to possible damping at the structure
vertex for virtual momentum near to $m_L$, where the integral gets its main
contribution. At present, the significance of a possible experimental deviation\cite{ref7}
from expectation is reduced by the correction of a long-standing sign error in the
calculation of a ``standard'' theoretical contribution.\cite{ref8,ref9} In this paper,
we consider explicitly, only neutrino, effective neutral-current couplings. 
}
The essential physical idea is that of neutrino structure at very small distances
($<10^{-18}\cm$), due to hadron-like coupling to $L$. For this $m_L$, $E_\nm$ must
reach $\sim 2\times 10^{21}\eV$ for production of real $L$. However, the
amplitude for the diagram in Fig.~(1), which involves a virtual $L$, results
in a cross section above the electroweak $\sigma_\nm\sim 10^{-31}\cm^2$, already at
$E_\nm\sim 10^{20}\eV$.\cite{ref5} For a somewhat smaller $m_L\sim 0.3\times 10^6\GeV$,
the production of real $L$ is possible for $E_\nm$ above about $0.5\times 10^{20}\eV$, which
is close to where the GZK cut-off\cite{ref3,ref4} is located\cite{ref1,ref2} for
protons coming from cosmological distances. In this case $\sigma_\nm$ can be of the order
of millibarns at $E_\nm\sim 10^{20}\eV$. Then, a dip at the GZK cut-off, followed by a
``bump-structure''\cite{ref10} near to $10^{20}\eV$ could occur in the air-shower
energy spectrum, if some of the highest-energy events are induced by the primary neutrinos
interacting in the atmosphere.\footnote{
There must be a sufficient flux of neutrinos near to $10^{20}\eV$, say $>10^{-19}\cm^{-2}\s^{-1}\sr^{-1}$.
A possible origin is from the two-body decay of a form of massive dark matter, with
a lifetime ($\propto m_{\nu_\tau}^{-2}$)which is many orders of magnitude greater than the present
age of the universe.\cite{ref10} Neutrino-initiated air showers do not necessarily fall off
strongly at larger zenith angles. (An indirect, dynamical mechanism which involves
a sizable flux of very high-energy neutrinos as the origin of an air-shower ``burst'',
postulates resonant interaction with CMB neutrinos (for non-zero mass), through a 
light $Z'$\cite{ref13}, as in ref.~14 for $Z$.) 
} Such a spectrum behavior is consistent
with present data.\cite{ref1,ref2} This may only reflect insufficient exposure to
obtain events at much higher energies. On-going and new cosmic-ray experiments will tell.\cite{ref2} 

The purpose of this paper is to show that the kind of hypothetical neutrino structure
summarized above, can have an effect upon the extraction of $\st$ from muon neutrino-hadron
neutral-current cross section data at $E_\nu < 300\GeV$. The motivation for this is
the recent NuTeV result\cite{ref11} which indicates $\st=0.2276\pm 0.0017$, about
$2\%$ higher than the value deduced mainly from the precision data at LEP,
$\st = 0.2226\pm 0.0004$.\cite{ref12} The NuTeV value of $\st$ corresponds to a
$W$ mass about 300 MeV below the direct measurements at LEP II and at the Tevatron.\cite{ref12}
It may well be that the NuTeV value of an effective $\st$ relevant to $\nm$-hadron scattering
is to be explained in terms of violation of isospin symmetry in the parton distributions
of the nucleon, or in terms of an asymmetry in the strange quark sea.\cite{ref13} It is 
common to speculate\cite{ref15} about exchange of higher mass $Z'$ bosons (usually
in the TeV range), whose amplitude interferes with  that from $Z$ exchange.
The new amplitude must be such as to give destructive interference. This reduces
the neutral-current cross section.\footnote{
This is the likely direction implied by the experimental effect in the measured
ratio $(\sigma_\nm^{\mathrm{NC}}/\sigma_\nm^{\mathrm{CC}})\sim 1/2-\st + (5/18)(\st)^2$,
(given the $\sim 2\%$ precision in the measurements\cite{ref16} of $\sigma_\nm^{\mathrm{CC}}$).
} In a particular case\cite{ref13},
the $Z'$ coupling is required to conserve parity (i.~e.~no axial vector coupling).
This is not the only way to obtain a parity-conserving, effective four-fermion
interaction for $\nm$-hadron neutral-current scattering. A short-distance neutrino
structure, of the kind discussed above, can also be characterized by such a parity-conserving,
effective interaction, and can involve a very high particle mass. Our result below, suggests
that for $m_L\sim 0.3\times 10^6\GeV$ an effective $\st$ obtained from analysis of
NuTeV data, might be up to $\sim 0.005$ larger than the expected electroweak value.
Thus, a quantitative connection is made to the possibility of stronger neutrino interactions
at the highest energies.

In Fig.~(2), we illustrate schematically the reduction of the hypothetical primary
dynamics at very high $\sqrt{s}$ approaching $m_L$, to an approximate, effective
four-fermion interaction relevant for the NuTeV $\nm$-nucleon scattering data
at $E_\nm < 300\GeV$. The essential assumption is the following. The calculation
of a $\nm$ cross section\cite{ref5} for the hypothetical process in Fig.~(1), involves using a
typical hadronic total cross section to give an approximate strength for the sum
of processes included in the square of the lower ``vertex''. This is a controlling
factor in the size of the $\nm$ cross section. On the other hand,
interference of the standard-model amplitude from Fig.~(3), with an effective
amplitude from Figs.~(2), requires a degree of coherence for processes within
the left-hand ``black box'' in Fig.~(2a).\footnote{
In Figs.~(2,3), the box on the right represents hadronic fragmentation of partons, after
the basic partonic collision process.
} We assume that this
is achieved as illustrated in Fig.~(2b). The axial-vector $A^0$ related 
to the $\pi^0$ in Fig.~(1), propagates from the hypothetical (undamped) $\nm-L$
vertex to interaction with the target nucleon via a hard parton-parton collision;
the result is a scattered nucleonic parton and an isoscalar, scalar ``jet''\footnote{
The jet is in effect, a neutral pair of pions, in the present context of hypothetical
(point-like) $\pi^0$, $A^0$ interactions at the $\nm-L$ effective vertices.
}
(denoted by the exchanged double line ($j^0$) in Figs.~(2b,c)). In Fig.~(2c),
two vertices (with strength magnitude $|gg'|\sim g^2$) and the intermediate
state on the neutrino line, characterized by the very high mass $m_L$,
are brought together to an effective form $(g^2/m_L)(\ol{\nu}\gamma_\mu\gamma_5\nu)$.
Similarly for the (coherent) intermediate states in the partonic ``black box''
on the quark line in Fig.~(2b), with an assumed real part for the effective vertex.
An effective Lagrangian can be approximately represented by the form
\be
{\cal L}_{\mathrm{eff}} = -\left( \frac{\sqrt{2}g^2}{m_L m}\right)\left(
\ol{\nm}\gamma_\mu\gamma_5\nm\right)\left(\ol{q}\gamma^\mu\gamma_5\tau_3 q\right)
\ee
where $q$ represents a $u$ or $d$ quark, and $m$ is an effective mass given by the product
of significant numerical factors of $\pi$ (see Eq.~(4) below) and a dynamical mass
$\sim 3\GeV$ which characterizes the ``black box''. The negative sign before 
${\cal{L}}_{\mathrm{eff}}$ ( for $q=u$) corresponds to destructive interference
of the amplitude with the standard-model amplitude from Fig.~(3). The negative
sign can arise from the structure of the intermediate states in Fig.~(2)
and the overall sign of the product of effective vertices. This sign appears
to arise naturally when the hadronic black box in Fig.~(2b) involves a hard parton
scattering (gluon exchange). The order of magnitude of the ratio of the amplitude
from Fig.~(2c) to that from Fig.~(3) is
\be
|R|\sim\left(\frac{g^2}{e^2}\right)\left(\frac{m_Z}{m_L}\right)\left(\frac{m_Z}{m}\right)
\sim 0.002
\ee
for $g^2/(4\pi) \sim 1$, $m_L \sim 0.3\times 10^6 \GeV$,
$e^2/(4\pi)\cong \frac{1}{137}$, $ m_Z \cong 90 \GeV$ and with
$m\sim(20\pi^3)(3\GeV)\sim 1.8 \times 10^3\GeV$ (calculated in Eq.~(4) below).

In Eq.~(1), the effective strength is
\be
\left(\frac{\sqrt{2}g^2}{m_Lm}\right)\sim 0.33\times 10^{-7}\approx \left(\frac{G_F}{\sqrt{2}}\right)
\frac{(0.01)}{2}
\ee
The value $\sim (G_F/\sqrt{2})(0.01)$ has been deduced \cite{ref13} phenomenologically as
the approximate necessary strength of an effective four-fermion interaction which is, a priori,
required to be $SU(2)_L$ invariant (we use only a corresponding parity-conserving, axial-axial
term involving $\ol{\nu}\nu$ and $\ol{u}u$(-$\ol{d}d$)). In the present model, the effective
strength in Eq.~(3) is directly fixed by the hypothesis of relatively strong, parity-conserving
neutrino interactions (thus $g^2/4\pi \sim 1 $ compared to $e^2/4\pi$), at $E_\nu\sim 10^{20}\eV$
(corresponding to $m_L\simlt 10^6\GeV$). The effective mass $m$ is determined by the dynamical
structure of the model, as we now estimate. The neutrino cross section (here assumed to be
relevant for p or n in an isoscalar target), calculated
approximately from the amplitude in Fig.~(2b), is
\ba
\Delta \sigma_\nm &\sim& (g^2)^2\left\{ \left(\frac{1.22}{(4\pi)^2}\right)^2
\left(\frac{\langle n_j \rangle \sigma_j}{2\pi^2}\right)\right\}\left\{
\left( \frac{4}{3\pi}\right)\left(\frac{s}{4m_L^2}\right)\right\}\nn\\
   &=&(g^2)^2\left(\frac{1}{m^2}\right)\left\{\left( \frac{4}{3\pi}\right)\left(\frac{s}{4m_L^2}\right)\right\}
\cong (8\pi)^2\left\{ \frac{1}{(1.8\TeV)^2}\right\}(4.7\times 10^{-10})\nn\\
&\cong& (0.78\times 10^{-31}\cm^2)(4.7\times 10^{-10})\cong 0.37\times 10^{-40}\cm^2
\ea
In order to make clear the origin of the magnitude in this dynamical model, three
separate factors are enumerated in detail in Eq.~(4). There is the overall
strong strength from the two interactions on the neutrino  line, represented
by the magnitude $|gg'|\sim g^2$ ($g^2/4\pi \sim 2$, is used, as in ref.~5).
The second factor in curly brackets ($\{\ldots\}$) is kinematical, essentially
the phase space. At $\sqrt{s}\sim 20\GeV$, used in Eq.~(4) for illustration,
this is a small number of the order of $10^{-9}$. This number is controlled
by the large mass $m_L \sim 0.3\times 10^6\GeV$, whose size is approximately
fixed by the possibility of strong neutrino interactions at extremely
high energy, $\sim 10^{11}\GeV$. The first factor in curly brackets represents
the (inverse) square of an effective mass $m\sim (20\pi^3)(3\GeV)$; this
controls the effective strength of the hadronic ``vertex''. The mass has 
a large kinematic factor ($\sim 20\pi^3$) which is related to the specific
dynamics, i.~e.~to the integration over virtual momenta, and 
to the approximate parameterization of the ``black box''
in Fig.~(2b),  in terms of a measured cross section for a hard scattering
process. For the latter we use a typical jet cross section at relatively\cite{ref17}
low $\sqrt{s}$, $\sigma_j\sim 2\times 10^{-29}\cm^2$ in a central rapidity
interval $|\Delta y|\sim (1-(-1))=2$. We use this as a (fixed) parameter
{\underline{only in a limited}} range of the intergration momentum\footnote{
Note that a corresponding typical cross section\cite{ref18} for $\pi^0$, at
low $\sqrt{s}$ in this limited range for transverse momenta, is comparable,
about $0.4\times 10^{-30}\cm^2$. It falls away more rapidly than does $\sigma_j$,
at higher transverse momenta.
},
$\sim 2\GeV$ to $\sim 3.7\GeV$; $\sigma_j$ falls away at high loop momentum.\footnote{
The approximate integration over virtual momenta $Q$, with $\sigma_j$ fixed over
a limited range, involves a factor $\ln(Q^2_{\mathrm{max}}/Q^2_{\mathrm{min}})$.
This is of order unity; we use $2\ln(3.7/2)\sim 1.22$. If taken as unity, 
the interference term in Eq.~(5) has magnitude $\sim 0.0047$.
}
With the jet multiplicity $\langle n_j \rangle\cong 2 $, the quantity
$(\langle n_j\rangle \sigma_j)^{-1/2}$ translates into a mass of $\sim 3\GeV$,
which can be considered as the intrinsic, mass parameter characteristic of the
``black box'' in Fig.~(2b).

Using the estimate for $\Delta \sigma_\nm$ in Eq.~(4), and the empirical $\nu_\mu$ charged-current
cross section\cite{ref16}, $\sigma_\nm^{\mathrm{CC}}(E_\nm\sim 200 \GeV)\cong 1.36\times
10^{-36}\cm^2$, we calculate the magnitude of the interference term between
the amplitudes from Figs.~(2,3), in ratio to $\sigma_\nm^{\mathrm{CC}}$,
\be
|I|\cong 2\sqrt{\frac{\Delta \sigma_\nm}{\sigma_\nm^{\mathrm{CC}}}}\sqrt
{\frac{\sigma_\nm^{\mathrm{NC}}}{\sigma_\nm^{\mathrm{CC}}}}\cong
2(0.0052)(0.55)=0.0057
\ee
This is approximately independent of the specific $E_\nm$ used above
for illustrating the numbers. In Eq.~(5), we use $\sim 0.29$ for the ratio
of neutral to charged-current cross sections (i.~e.~as calculated$^{F3}$ for
$(\st)_{\mathrm{LEP}}=0.223$). A negative interference term of this magnitude
corresponds to an effective value of $\st$ in the NuTeV experiment which is
about 0.005 larger than the (electroweak) value determined from the LEP
measurements.

From  $\ol{p}p\to l^+l^-+X$ interactions, CDF has set limits\cite{ref19}
on quantities analogous to $\epsilon\sim 0.01$ in Eq.~(3), but referring
to effective four-fermion interactions involving quarks and a $\mu^-\mu^+$
(or $e^-e^+$) pair, (approximately, the $|\epsilon|$ are\cite{ref13}$\simlt 0.04$).
It is noteworthy that the hypothetical, specific $\nm$ structure in Fig.~(2)
does not have a counterpart for $\mu^-$.
This is because the scalar ``jet'' is neutral$^{F5}$. Thus, the experimental
limits\cite{ref19} from $q\ol{q}\to\mu^-\mu^+$ are not directly relevant here
(as they are for phenomenological interaction forms utilized in ref.~13).

There is a potentially interesting ``charge'' radius effect for $\nm$, implied
by the hypothetical structure. If $\rho^0$-exchange can occur in place of the
$A^0$ in Fig.~(2c), a charge radius is generated by replacing the lower
vertex by that for $\gamma\to \rho^0\pi^+\pi^-$, with a magnitude estimated as
\be
|\langle r^2\rangle_\nm|\sim \left(\frac{g^2}{64\pi^2}\right)\left(\frac{1}{\mu m_L}\right)
\sim 1.6\times 10^{-35}\cm^2
\ee
The mass parameter $\mu\sim 264\MeV$, is estimated from the measured partial width\cite{ref16}
$\Gamma(\rho^0\to \pi^+\pi^-\gamma)\sim 1.5\MeV$. Empirically\cite{ref16}, $|\langle r^2\rangle_\nm|
\simlt 10^{-32}\cm^2$. More generally, this represents a limit on certain non-standard
contributions to $\nm$ scattering\cite{ref16}; $\Delta \sigma_\nm$ in Eq.~(4) is much less.
At the Z-pole, a similar ``charge'' radius can affect, in principle, the magnitude of the axial-vector
(and possibly, vector) effective coupling to $\nm\ol{\nm}$. 
We have, $m^2_Z|\langle r^2\rangle_\nm|\sim 0.03\%$, which corresponds to a change
in this partial width of about $\pm 0.1\MeV$.
This means that the effective number of light neutrinos can be less than three\cite{ref15}.
There is $\sim 3\%$ accuracy
in the measurement\cite{ref20} of the $\nm$-coupling to $Z$ from $\nm$-electron scattering.
The direct measurement\cite{ref21} of the invisible decay width of the $Z$, is also
with $\sim 3\%$ accuracy. There is a fit value\cite{ref16} with $\sim 0.3\%$ uncertainty.

We summarize the model, the essential hypothesis for which involves the neutrino
production by effectively strong interactions, of a very massive neutral lepton at
the highest energies. The dynamics involves hypothetical, strong interactions of 
point-like components of exchanged hadronic entities which couple a neutrino to the
heavy lepton $L$; thus, a corresponding effective neutrino substructure at very 
short distances. The resulting calculated cross section\cite{ref5} can reach
above the millibarn level at $E_\nu\sim 10^{20}\eV$. The enhanced cross section
occurs already below threshold\cite{ref5}, and increases through the threshold
for production of real $L$. Clearly, the exchange mechanism in Fig.~1 involves
all angular momenta (this is not a unitarity-bounded, S-wave effect, as illustrated
in \cite{ref23}). In accord with estimates in \cite{ref23}, the scale (and dimension)
of the new cross section is set by a hadronic total cross section (from the 
square of the lower ``vertex'' in Fig.~1; as given explicitly in Eq.~(5) in \cite{ref5}).
In this paper, we have ``extrapolated'' from the hypothetical high-energy dynamics, to
a small effect at NuTeV energies. This arises from dynamical and kinematical
factors which are estimated numerically, and which are explicitly enumerated and
explained in the context of the high-energy model, following Eq.~(4) above.
At NuTeV $\sqrt{s}$, the relevant effective mass parameter is calculated to be about
4.7 TeV.

In conclusion, we note that while the present NuTeV effect might eventually get an
explanation in terms of asymmetries in parton distributions\cite{ref13}, ideas concerning
hypothetical new particles\cite{ref15,ref13} are not necessarily restricted to masses
below a few TeV. In the effective four-fermion interaction generated by structure,
in Eq.~(1), while the overall effective mass parameter is estimated as $\sim \sqrt{mm_L/g^2}
\sim 4.7 \TeV$, the particle mass $m_L\sim 300 \TeV$, is much higher. This is because
the hypothesis is made of a dynamical connection between the possibility of stronger
neutrino interaction in the atmosphere, reaching of the order of millibarn cross sections at the
highest energies $\sim 10^{11}\GeV$, and a small effect upon $\nm$-hadron scattering below
a few hundred GeV. This relates two presently puzzling phenomena: the cosmic-ray
air showers at $10^{20}$ eV and the possible anomalous deviation in the $\st$ extracted
from NuTeV data.\footnote{A limit on the possibility of a small amplitude arising from 
a hypothetical, point-like component of the pion coupling, to vector bosons\cite{ref22},
is discussed in the conclusion of\cite{ref5}. Possible observational consequences are:
(1) a partial width for $Z^0\to \pi^+\pi^-$ of the order of an MeV, and (2) a quite
small, but negative value, for the $S$ parameter\cite{ref15}, corresponding to a downward shift
in the $Z$ mass of the order of MeV.}

\section*{Added note}
After completion of this work, authors have called our attention to recent work on
large cross sections from brane theory; we include some references\cite{ref24,ref25,ref26}.
See also \cite{ref27} and \cite{ref28}, and the earliest related work \cite{ref29} which
we are aware of.

\newpage
\section*{Figures}
\begin{figure}[h]
\begin{center}
\mbox{\epsfysize 8cm\epsffile{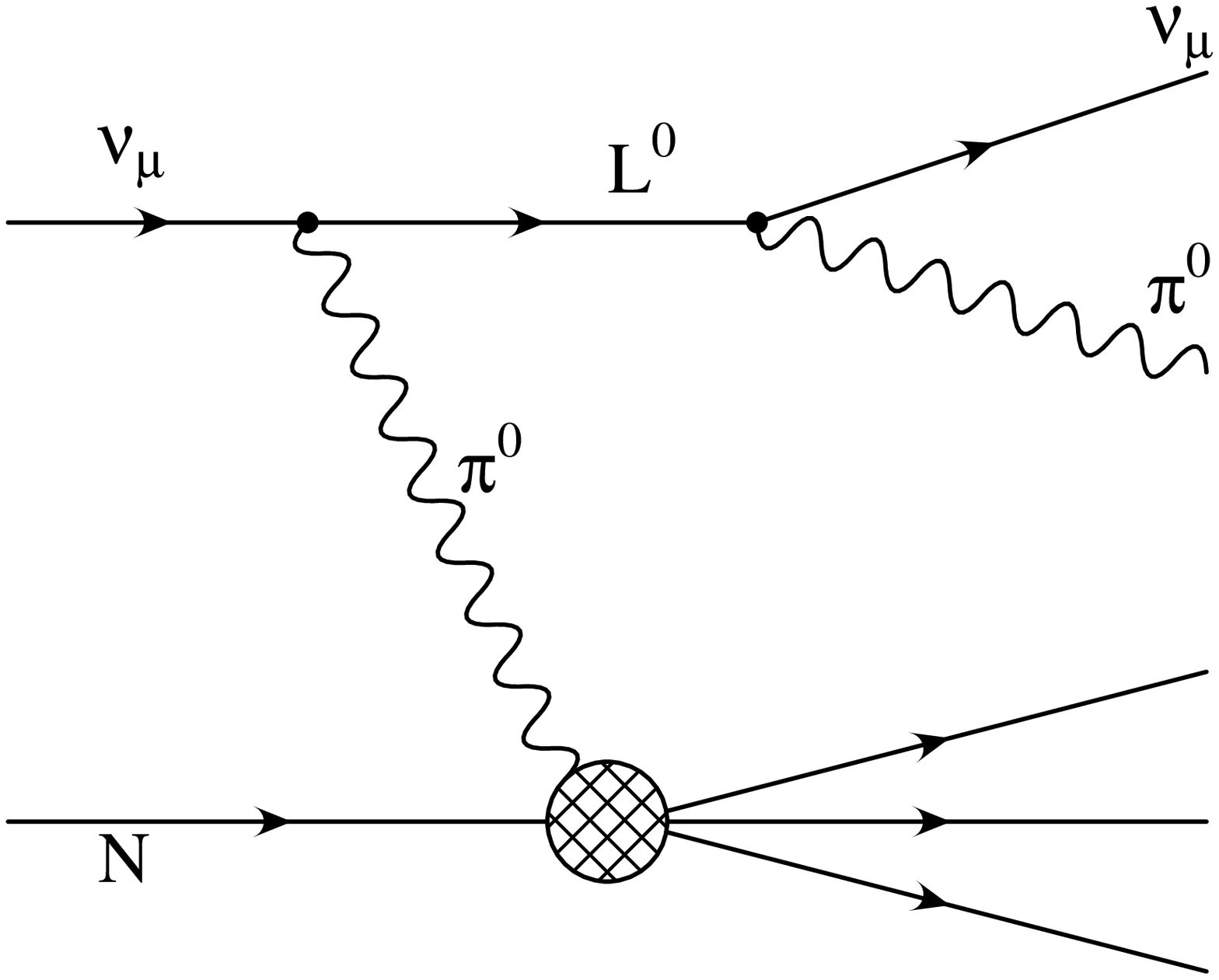}}
\caption{Feynman diagram for a contribution to the interaction cross
section of an extremely high energy, cosmic-ray neutrino with an
atmospheric nucleon, mediated by a virtual $L^0$.}
\end{center}
\end{figure}
\begin{figure}[b]
\begin{center}
\mbox{\epsfysize 13cm\epsffile{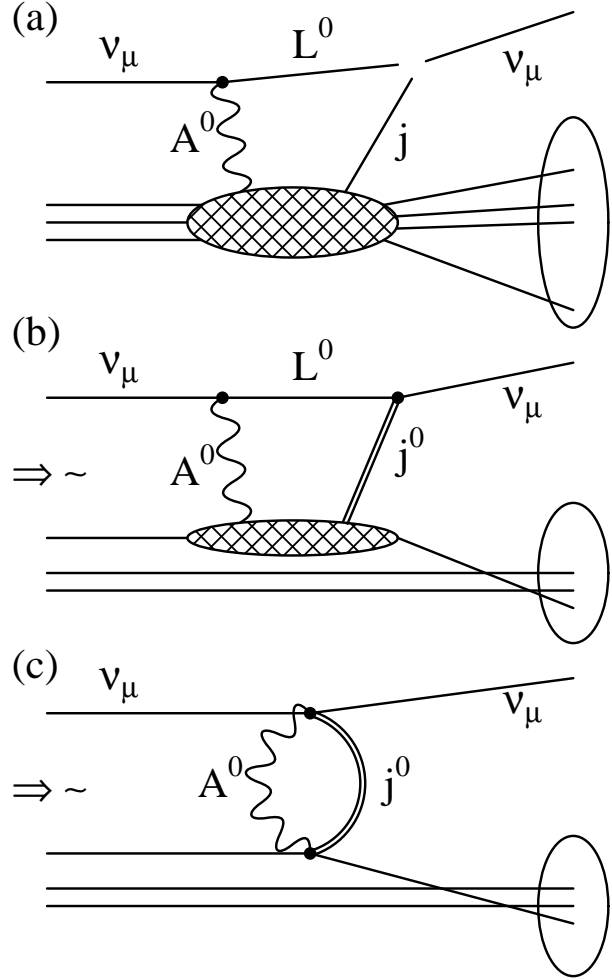}}
\caption{For $\nm$-hadron scattering at $\sqrt{s}\sim 20\GeV$, a schematic
sequence leading to the approximate effective four-fermion interaction in Eq.~(1), as
described in the text and footnotes F4-7. The effective vertices on the neutrino
line are assumed to be point-like, up to momenta near to the scale of $m_L$.
The loop integration momentum is taken as relevant for jet production, but
only a limited range is used in estimating the amplitude for Eq.~(4),
 $2\GeV\simlt |Q|\simlt 3.7\GeV$, because $\sigma_j$ falls away at high $|Q|$.}
\end{center}
\end{figure}
\begin{figure}[t]
\begin{center}
\mbox{\epsfysize 8cm\epsffile{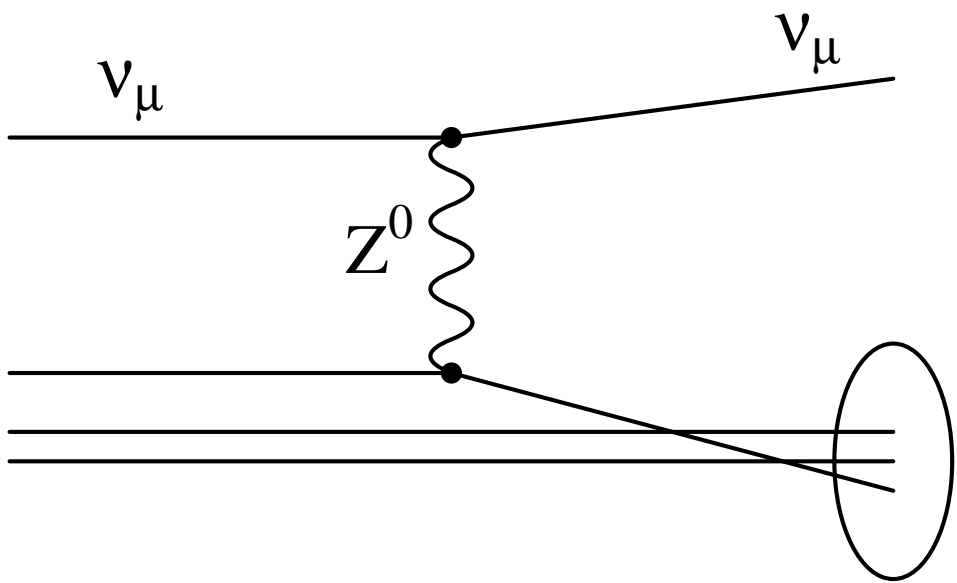}}
\caption{The standard model interaction.}
\end{center}
\end{figure}

\begin{thebibliography}{99} 
\bibitem{ref1} M.~Takeda et al., Phys.~Rev.~Lett.~{\bf 81}, (1998) 1163 
\bibitem{ref2} M.~Nagano and A.~A.~Watson, Rev.~Mod.~Phys.~{\bf 72} (2000) 689
\bibitem{ref3} K.~Greisen, Phys.~Rev.~Lett.~{\bf 16} (1966) 748
\bibitem{ref4} G.~T.~Zatsepin nad V.~Kuzmin, JETP Lett.~{\bf 4} (1966) 78
\bibitem{ref5} S.~Barshay and G.~Kreyerhoff, Eur.~Phys.~J.~{\bf C23} (2002) 191
\bibitem{ref6} B.~E.~Lautrup, A.~Peterman and E.~de Rafael, 
Phys.~Rep.~{\bf 3C} (1972) 193
\bibitem{ref7}Muon $(g_\mu-2)$ Collab., H.~M.~Brown et al., Phys.~Rev.~Lett.~{\bf 86} (2001)
2227
\bibitem{ref8} M.~Hayakawa and T.~Kinoshita, hep-ph/0112102
\bibitem{ref9} M.~Knecht and A.~Nyffeler, hep-ph/0111058
\bibitem{ref10}S.~Barshay and G.~Kreyerhoff, Nuovo Cimento {\bf 112A} (1999)
1463; {\bf 112A} (1999) 1469 
\bibitem{ref11}G.~P.~Zeller et al., hep-ex/0110059v2
\bibitem{ref12}J.~Drees, hep-ex/0110077v1
\bibitem{ref13}S.~Davidson et al., hep-ph/0112302v2
\bibitem{ref14}T.~Weiler, Phys.~Rev.~Lett.~{\bf 49} (1982) 234 and erratum
ibidem {\bf 12} (2000) 379
\bibitem{ref15} J.~Erler and P.~Langacker, Phys.~Lett.~{\bf B456} (1999) 68\\
J.~Erler and P.~Langacker, Phys.~Rev.~Lett.~{\bf 84} (2000) 212\\
P.~Langacker, hep-ph/0102085
\bibitem{ref16} Particle Data Group, Rev.~Part.~Phys., Eur.~Phys.~J.~{\bf C15} (2000) 379
\bibitem{ref17} AFS Collab., T.~Akesson et al., Phys.~Lett.~{\bf 123B} (1983) 133
\bibitem{ref18} C.~Kourkoumelis et al., Z.~Phys.~{\bf C5} (1980) 95
\bibitem{ref19} CDF Collab., F.~Abe et al., Phys.~Rev.~Lett.~{\bf 79} (1997) 2198
\bibitem{ref20} CHARM II Collab., P.~Vilain et al., Phys.~Lett.~{\bf B320} (1994) 203
\bibitem{ref21} L3 Collab., M.~Acciarri et al., Phys.~Lett.~{\bf 431} (1998) 199
\bibitem{ref22} S.~Barshay, Phys.~Lett.~{\bf B286} (1992) 375
\bibitem{ref23} G.~Burdman, F.~Halzen, and R.~Gandhi, Phys.~Lett.~{\bf B417} (1997) 107
\bibitem{ref24} L.~A.~Anchordoqui, J.~L.~Feng, and H.~Goldberg, hep-ph/0202124
\bibitem{ref25} E-J. Ahn, M.~Cavagli\`a, and A.~V.~Olinto, hep-th/0201042
\bibitem{ref26} P.~Jain, S.~Kar, S.~Panda, and J.~P.~Ralston, hep-ph/0201232
\bibitem{ref27} G.~Domokos and S.~Kovesi-Domokos, Phys.~Rev.~Lett.~{\bf 82} (1999) 1366
\bibitem{ref28} J.~Bordes et al., Astropart.~Phys.~{\bf 8} (1998) 135
\bibitem{ref29} V.~Berezinsky and G.~Zatsepin, Phys.~Lett.~{\bf B28}, (1969) 423
\end{thebibliography}
\end{document}